\documentclass[aip,apl,graphicx,reprint]{revtex4-1}
\usepackage{graphicx}
\usepackage{color,soul}

\begin{document}

\title{Perpendicular magnetic anisotropy in Co$_2$MnGa} 

\author{B.\,M.\,Ludbrook}
\author{B.\,J.\,Ruck}

\affiliation{The MacDiarmid Institute for Advanced Materials and Nanotechnology, School of Chemical and Physical Sciences, Victoria University of Wellington, P.O. Box 600, Wellington 6140, New Zealand}

\author{S.\,Granville}
\affiliation{The MacDiarmid Institute for Advanced Materials and Nanotechnology, Robinson Research Institute, Victoria University of Wellington, P.O. Box 33436, Lower Hutt 5046, New Zealand}

\date{\today}

\begin{abstract}
We report perpendicular magnetic anisotropy in the ferromagnetic Heusler alloy Co$_2$MnGa in a MgO/Co$_2$MnGa/Pd trilayer stack for Co$_2$MnGa thicknesses up to 3.5 nm. There is a thickness- and temperature-dependent spin reorientation transition from perpendicular to in-plane magnetic anisotropy which we study through the anomalous Hall effect. From the temperature dependence of the anomalous Hall effect, we observe the expected scaling of $\rho_{xy}^{AHE}$ with $\rho_{xx}$, suggesting the intrinsic and side-jump mechanisms are largely responsible for the anomalous Hall effect in this material.

\end{abstract}

\pacs{}

\maketitle 


The cobalt-based Heusler alloys are coming under intense scrutiny due to the half-metallic nature of their electronic structure, and their ferromagnetism with high Curie temperature and low Gilbert damping parameter.\cite{Kubota2009} In particular, Co$_2$FeSi and Co$_2$MnSi both have 100\% spin polarization\cite{Kandpal2006,Yang2013,Jourdan2014} and high Curie temperatures ($\approx$ 800$^{\circ}$C). These materials show great promise for applications in spin-transfer-torque and magnetoresistive devices.

The less studied Heusler compound Co$_2$MnGa (CMG) is also ferromagnetic with a T$_C$ of 700 K, and a relatively low saturation magnetization of 700 emu/cm$^3$.\cite{Pechan2005} It is predicted to be a type III half-metal\cite{Felser2013} where the majority and minority spin electrons are itinerant and localized respectively.\cite{Coey2002} It has the important characteristic of a high resistance to oxidation, which makes it promising for devices. This has enabled surface sensitive studies of its electronic structure, which have found a 55\% spin-polarized density of states and good agreement with the calculated band structure.\cite{Hahn2011,Kolbe2012} It is also promising for lattice matched integration with GaAs semiconductors.\cite{Claydon2007} However, relatively little is known about the electronic and magnetic properties of this material in the sorts of multilayer stacks required for applications.

Perpendicular magnetic anisotropy (PMA) is an important property in a magnetic material, enabling applications such as low-power spin-transfer-torque devices. Early studies suggested Fe-O hybridization at an interface was the mechanism behind PMA in CoFeB\cite{Ikeda2010} and subsequent work with Heusler compounds focused on compounds containing Fe on MgO.\cite{Wen2011,Li2011a,Ding2012a,Cui2013,Okabayashi2013,Wen2014} PMA has recently been reported in Co$_2$MnSi and Co$_2$Fe$_{x}$Mn$_{1-x}$Si adjacent to a Pd layer,\cite{Kamada2014,Matsushita2015,Ludbrook2016} indicating the Fe-O hybridization is not the only way to obtain PMA in a magnetic thin film. Here, we demonstrate PMA in MgO/Co$_2$MnGa/Pd stacks, and show how a spin reorientation transition (SRT) between in-plane magnetic anisotropy and PMA occurs with the variation of both temperature and Co$_2$MnGa thickness. Understanding these properties in the Heuslers is an important step towards the implementation of robust spintronic devices using highly spin-polarized materials.

\begin{figure}[h!]
\includegraphics[width=8.5cm, hiresbb=true]{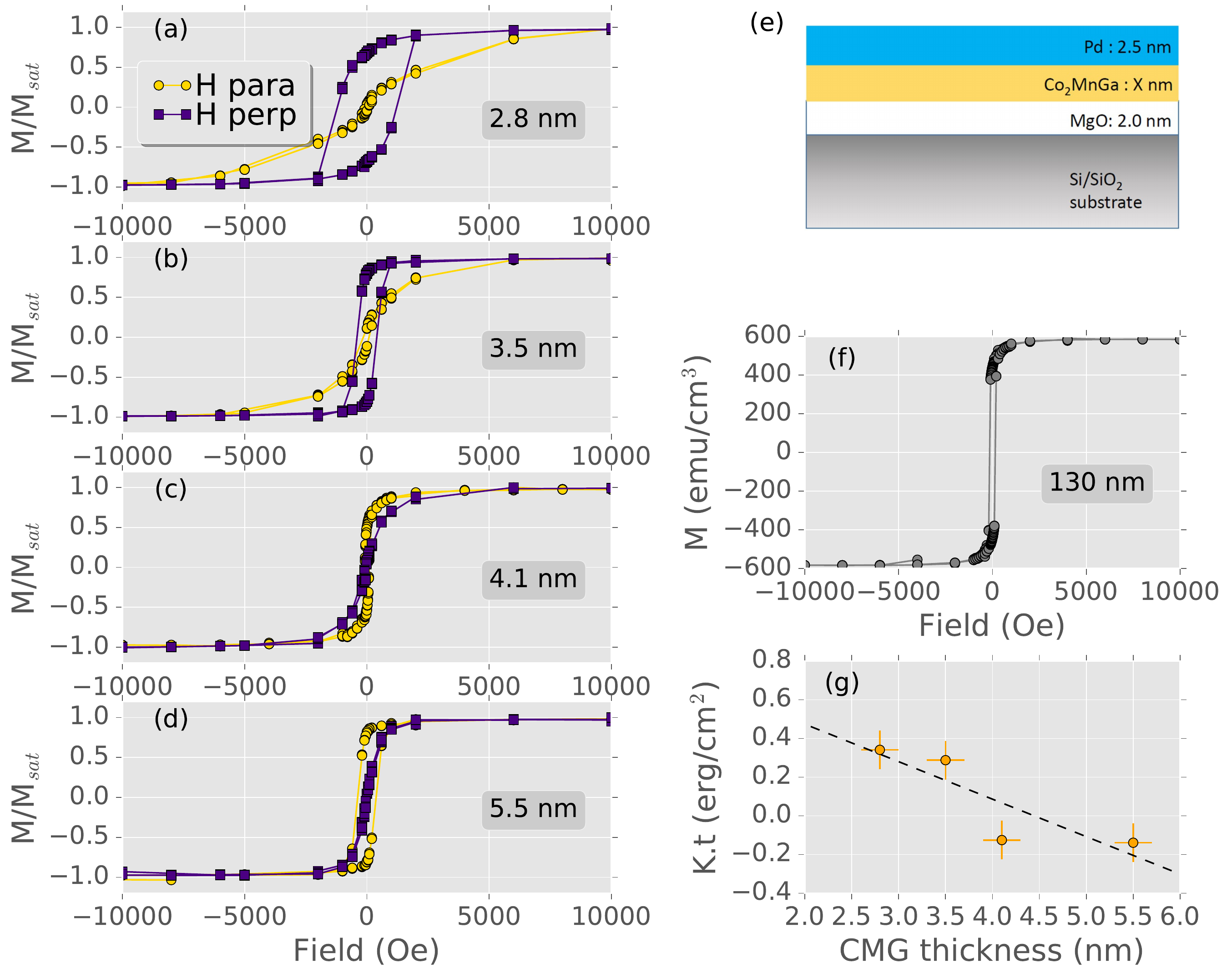}
\caption{\label{Fig:1} (a-d) Normalized magnetization versus field loops at 300\,K for the trilayer stacks measured with the field parallel to the film plane (gold circles) and perpendicular to the film plane (indigo squares). The thickness of the Co$_2$MnGa layer is given in each plot. The trilayer structure is sketched in (e). (f) Magnetization versus field at 300\,K for a 130 nm thick Co$_2$MnGa film measured with the field in-plane. (g) The magnetic anisotropy energy density extracted from the data in panels (a-d) is multiplied by the Co$_2$MnGa film thickness and plotted against the Co$_2$MnGa thickness to quantify the PMA strength. }
\end{figure}

Thin films were grown by DC magnetron sputtering in a Kurt J Lesker CMS-18 UHV system with a base pressure of $2\times 10^{-8}$ Torr. Multilayer stacks were prepared on $10 \times 10$mm Si (+ native oxide) substrates in the sequence MgO(2)/Co$_2$MnGa(t)/Pd(2.5), where the number in parentheses is the nominal layer thickness in nanometers. The trilayer structure is sketched in Fig.\,\ref{Fig:1}(e). Samples were grown at ambient temperature and post-growth annealed in-situ for 1 hour at 300$^{\circ}$C. MgO was RF sputtered at a growth rate of 0.05 \AA/s. Co$_2$MnGa and Pd were DC sputtered at growth rates of 0.69 \AA/s and 4.0 \AA/s respectively. Growth rates were calculated by growing a thick ($>50$~nm) film and measuring the thickness with a Dektak profilometer and Rutherford backscattering spectrometry. The composition of the Heusler target was verified to be Co$_2$MnGa by energy dispersive x-ray analysis in a SEM. Magnetization and Hall resistance measurements were done in a Quantum Design SQUID and PPMS respectively. The Hall measurements were made using sample-holders from Wimbush Science \& Technology, with spring-loaded contacts in a van der Pauw geometry.


Normalized magnetization versus field loops for the trilayer films shown in Figs.\,\ref{Fig:1}(a-d) demonstrate the PMA for Co$_2$MnGa thicknesses below 3.5 nm. This is evidenced by the high remnant magnetization when the magnetic field is applied perpendicular to the film plane (indigo squares in Figs.\,\ref{Fig:1}(a) and (b)) and the large field required to saturate the magnetization when the field is applied parallel to the film plane (gold circles). This behavior is inverted for the thicker Co$_2$MnGa layers in Figs.\,\ref{Fig:1}(c) and (d). A thicker (130 nm) Co$_2$MnGa film grown on MgO buffered Si and annealed under the same conditions as the trilayers has an in-plane easy axis of magnetization with a saturation moment of $600\pm 60\,\mathrm{emu/cm}^3$, shown in Fig.\,\ref{Fig:1}(f).

The magnetic anisotropy energy density (K$_U$) can be quantitatively determined from the area enclosed between the in-plane and out-of-plane data. We calculate this by taking the difference in the integrals of the data between 0 and $\pm 10000$ Oe. Positive values of $\int{[M_{\perp}-M_{\parallel}]}dH$ correspond to PMA, and negative values to in-plane anisotropy. We find a maximum $K_U = 1.3 \pm 0.2 \times 10^6\,\mathrm{erg/cm}^3$ for the 2.8 nm thick Co$_2$MnGa film. The uniaxial anisotropy energy density can be expressed as $K_U=K_V - 2 \pi M_s^2 + K_S/t$, where $K_V$ and $K_S$ are the volume and interface anisotropy terms respectively, and the term $2 \pi M_s^2$ is due to the shape anisotropy. Plotting $K_U\times t_{Co_2MnGa}$ against the film thickness in Fig.\,\ref{Fig:1}(g) allows one to separate the volume and interface contributions to K$_U$.\cite{Vaz2008a} The data lie on a straight line where the intercept gives $K_S=0.9 \pm 0.2\,\mathrm{erg/cm}^2$, implying it is the interface contribution that stabilizes PMA and dominates at low thicknesses. This behavior, and the values of $K_U$ and $K_S$ are in agreement with similar behavior reported in several other Heusler compounds recently.\cite{Wen2011,Ludbrook2016,Matsushita2015}


\begin{figure}[t!]
\includegraphics[width=8.5cm, hiresbb=true]{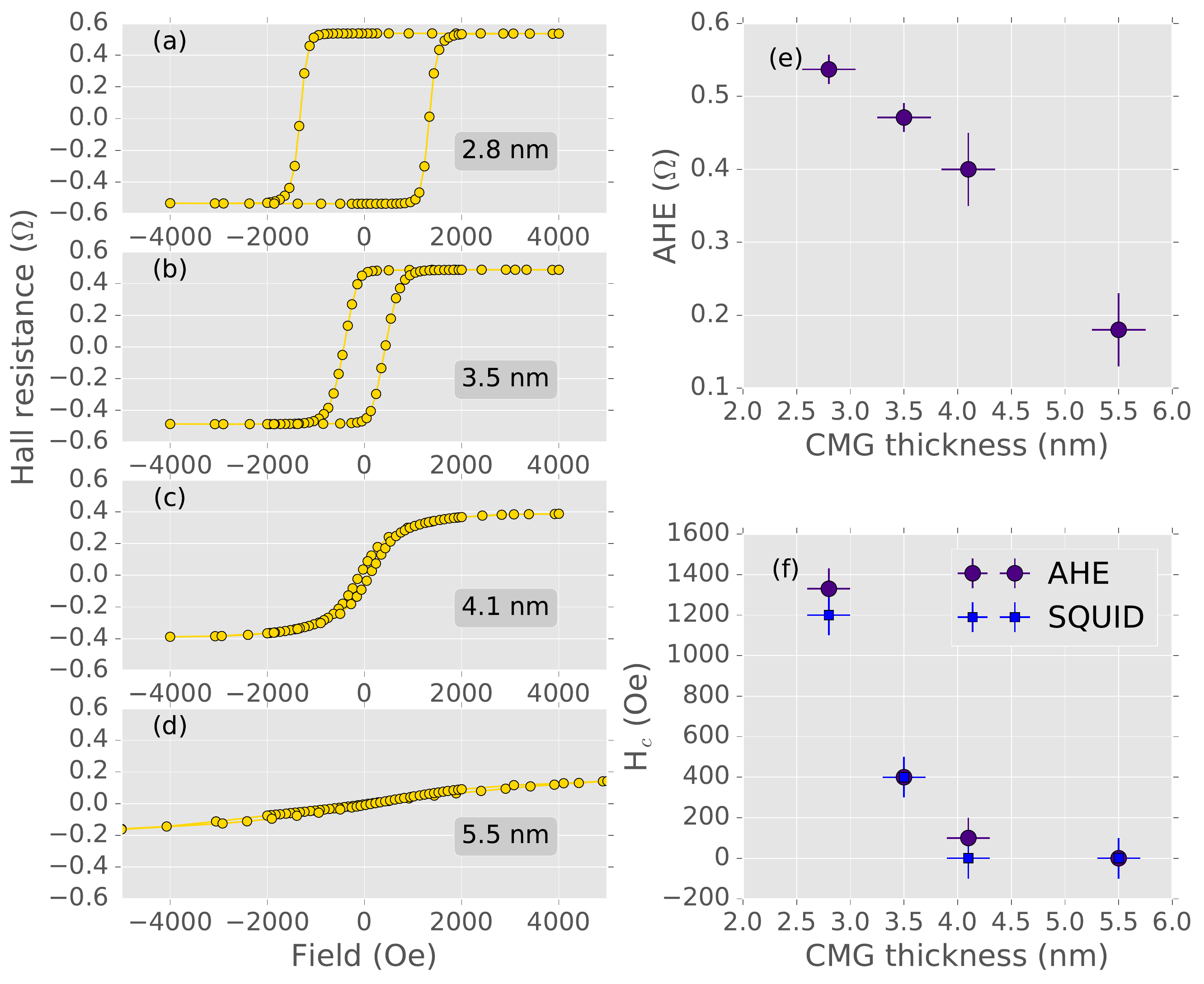}
\caption{\label{Fig:2} (a)-(d) Hall effect measurements at 300\~K on the  trilayer stacks with varying Co$_2$MnGa thickness show a strong anomalous Hall effect. The 100\% remnance confirms the PMA for the thinner films in (a) and (b). The magnitude of the AHE is determined by extrapolating the high-field Hall data back to zero field, and is shown in (e). The coercive field determined from the AHE (circles) and from the SQUID magnetization in perpendicular field (squares) is shown as a function of Co$_2$MnGa thickness in (f).}
\end{figure}

The Hall resistivity measured in a ferromagnetic material is empirically given by $\rho_{xy} = R_H H_z + R_S M_z$.\cite{Nagaosa2010} It is the sum of the ordinary Hall effect (OHE), linear in applied field ($H_z$), and the anomalous Hall effect (AHE), proportional to the out-of-plane magnetization ($M_z$). The anomalous Hall resistivity is determined by extrapolating the high-field data to zero-field. The AHE in the Hall loops in Figs.\,2(a)-(d) confirms the room temperature PMA for Co$_2$MnGa film thicknesses up to 3.5~nm, which show 100 \% remnance and a significant coercive field. The thicker films in Figs.\,\ref{Fig:2}(c) and (d) show an AHE characteristic of an in-plane easy magnetic axis, with a saturation field on the order of several kOe.

\begin{figure*}[t!]
\includegraphics[width=1\linewidth, hiresbb=true]{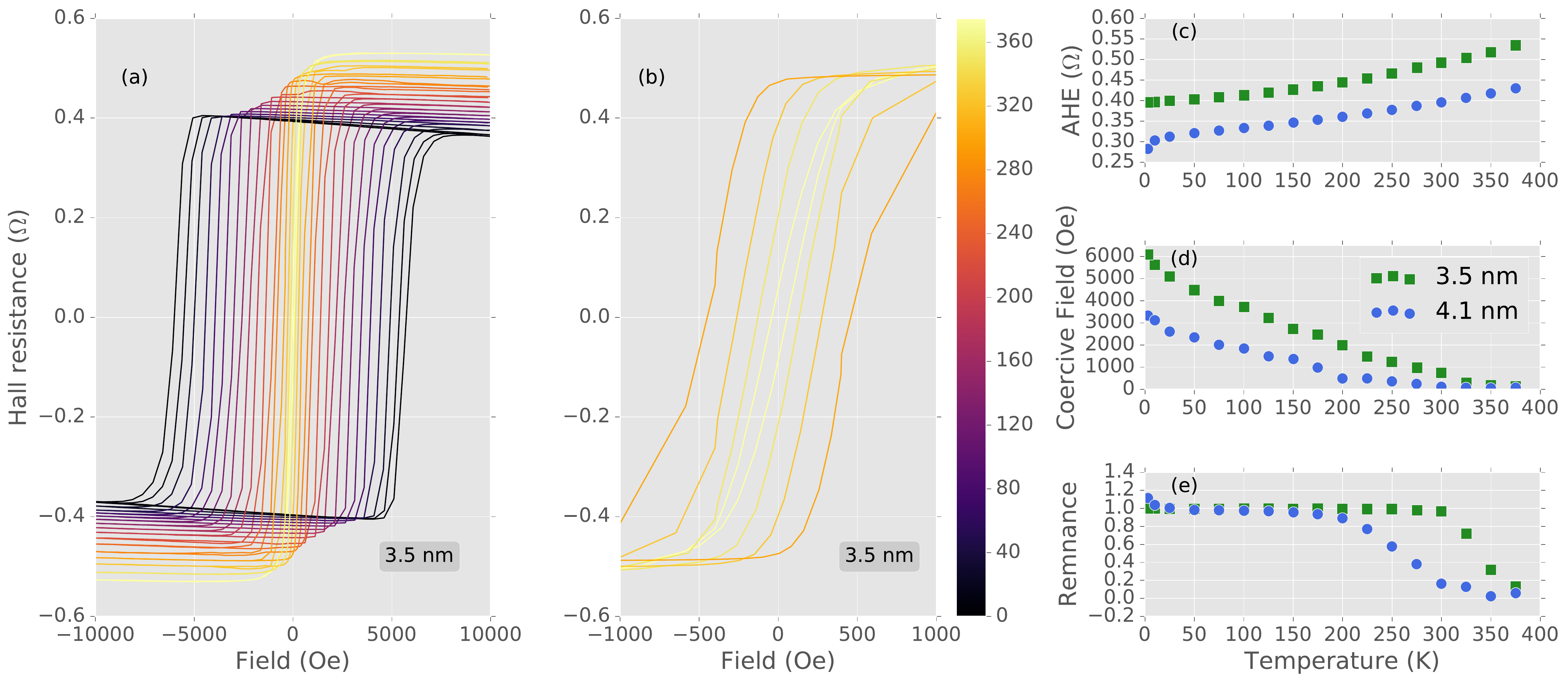}
\caption{\label{Fig:3} (a) The Hall effect measured at temperatures from 375 K to 3 K for a sample with $T_{Co_2MnGa} = 3.5 nm$. The temperature is given by the colorbar. Measurements at 300, 325, 350 and 375 K are shown over a narrower field range in (b), showing the transition from PMA to in-plane magnetic anisotropy. The AHE resistance and coercive field as a function of temperature are shown in (c) and (d) respectively for two Co$_2$MnGa thicknesses. The remnance, defined as the ratio of the zero-field magnetization to the saturation magnetization, is shown in (e).}
\end{figure*}

The magnitude of the AHE varies with film thickness, as shown in Fig.\,\ref{Fig:2}(e). This will be further discussed below in relation to the temperature dependent measurements. The coercive field determined from both the AHE and magnetization, shown in Fig.\,\ref{Fig:2}(f), also varies strongly with film thickness, in contrast to similar measurements on the Heusler compound Co$_2$Mn$_x$Fe$_{1-x}$Si where the room temperature coercive field was small ($\le 100 \mathrm{Oe}$) for all thicknesses.\cite{Ludbrook2016} In an ideal ferromagnet, the coercive field is given by $H_c=2K_U/M_s$. The saturation magnetization $M_s$ of Co$_2$MnGa is about 2/3 that of Co$_2$Mn$_x$Fe$_{1-x}$Si, which provides an explanation for the larger $H_c$ in Co$_2$MnGa. In fact, the ability to tune the coercive field with thickness is a useful property for device engineering, where hard and soft magnetic layers are needed.  


Hall sweeps measured from 375 K to 3 K are shown in Fig.\,\ref{Fig:3}(a) for the 3.5 nm thick Co$_2$MnGa film in the trilayer stack. The color corresponds to the temperature, as shown in the colorbar (in units K). The four measurements between 300 K and 375 K for the 3.5 nm thick Co$_2$MnGa layer in Fig.\,\ref{Fig:3}(b) show both the coercive field and the remnance going to zero as the temperature is increased. This indicates a spin reorientation transition (SRT) from PMA to in-plane magnetic anisotropy above 300 K.\cite{Pappas1990} Temperature dependent measurements on a  thicker 4.1 nm Co$_2$MnGa layer, which did not show PMA at room temperature, showed a SRT to PMA between 300 and 275 K. This temperature dependent SRT is seen clearly in the temperature dependence of the coercive field and remnance summarized in Figs.\,\ref{Fig:3}(d) and (e), where both quantities go to zero as the magnetization rotates into the film plane. The PMA in these stacks is controlled by the competition between the surface and volume anisotropy energy density. The temperature or thickness driven changes in the volume anisotropy term can explain the SRT between PMA and in-plane magnetic anisotropy.

\begin{figure}[b!]
\includegraphics[width=1\linewidth, hiresbb=true]{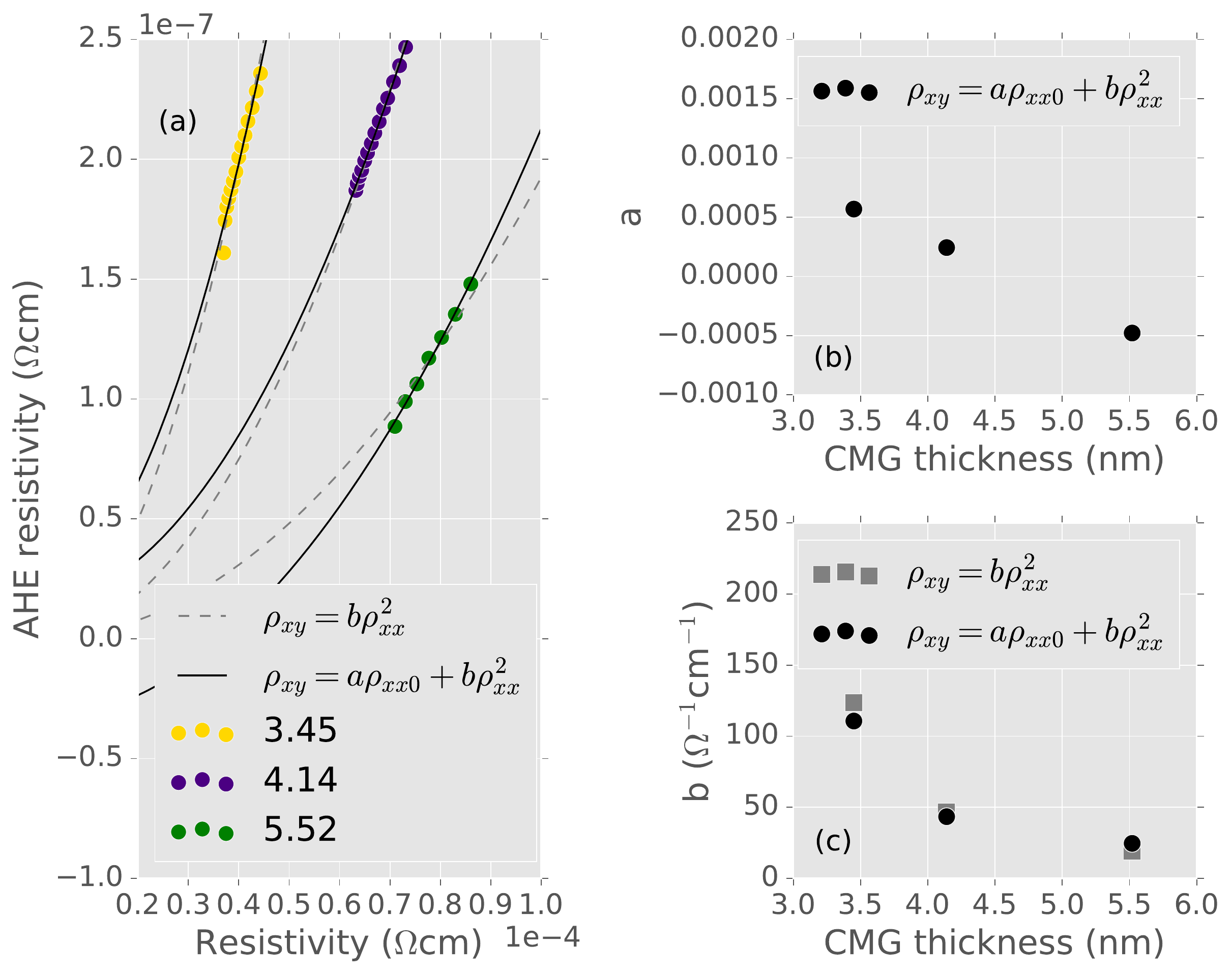}
\caption{\label{Fig:4} (a) From the temperature dependent data between 375 K and 30 K, the variation in $\rho_{xy}^{AHE}$ is plotted against $\rho_{xx}$ for each film thickness. The data is fitted with a pure quadratic term (dashed line) and a quadratic plus constant (solid line), described in the text. The fitting parameters $a$ and $b$ are shown in (b) and (c) respectively.}
\end{figure}

The magnitude of the AHE decreases with decreasing temperature, which can be seen in the raw data in Fig.\,\ref{Fig:3}(a), and plotted in Fig.\,\ref{Fig:3}(c). The AHE is proportional to the out of plane magnetization, and so the temperature dependence of the latter must be considered. SQUID measurements of a 130 nm Co$_2$MnGa film show a 6\% increase between room temperature and 3 K measured in a 130 nm Co$_2$MnGa film, in agreement with previous measurements.\cite{Finizio2015} The change in magnetization is small, but more importantly, it increases with decreasing temperature, and cannot explain the observed decrease in the AHE. Rather, variation in the AHE must be related to temperature-dependent changes in the skew- and side-jump- scattering rate, or changes in the Berry phase.\cite{Nagaosa2010} 

The various components of the AHE can be disambiguated by studying the scaling relationship between the anomalous Hall resistivity ($\rho_{xy}^{AHE}$) and the normal resistivity ($\rho_{xx}$). Although still not well understood, recent experimental and theoretical studies have suggested the scaling is described by\cite{Tian2009,Hou2015} $$ \rho_{xy}^{AHE} = a \rho_{xx0} + b \rho_{xx}^2 .$$ The first term corresponds to the skew scattering, while the second term relates to the intrinsic and side-jump terms. We use a two current model to account for the shunt current due to the Pd layer. The conductivity of a single Pd layer was measured as a function of temperature, and subtracted from the total conductivity of the multilayer films. Similarly, $\rho_{xy}^{AHE}$ is scaled by the ratio of the Co$_2$MnGa and Pd conductances to account for the reduced current flowing through the Co$_2$MnGa layer. In Fig.\,\ref{Fig:4}(a) the data and fits are shown for the above relation (solid line) as well as a pure quadratic fit (dashed line). The fitting parameters $a$ and $b$ are plotted in Figs.\,\ref{Fig:4}(b) and (c). The coefficient $b$ from the fits with only the quadratic term (i.e., $a=0$, plotted in gray) are not significantly different from the fit to the full equation (black points), indicating the skew-scattering term is of marginal significance. In fact, for the 3.45 nm film data, we find $a\rho_{xx0} = 0.036 \mu \Omega \mathrm{cm}$ and $b\rho_{xx0}^2 = 10 \mu \Omega \mathrm{cm}$, implying it is the intrinsic + side-jump term that dominates the AHE in this material.


We have shown how to induce PMA in the potentially useful Heusler alloy Co$_2$MnGa by sandwiching it between MgO and Pd thin films. The interface contribution to the magnetic anisotropy energy density is determined to be $K_S = 0.9 \pm 0.2 \,\mathrm{erg/cm}^2$ at ambient temperature, which favors PMA in Co$_2$MnGa layers with $t \leq$ 3.5~nm. The ferromagnetic Co$_2$MnGa undergoes a spin reorientation transition from perpendicular to in-plane magnetization with both increasing temperature and film thickness. The AHE has a temperature and thickness dependence that is not simply due to the magnetization, but rather can be traced back to variations in the skew- and side-jump- scattering, and intrinsic Berry phase contributions. It is the latter two that dominate in this material, although further work is required to further separate these two contributions. 

B.M.L. gratefully acknowledges post-doctoral funding from the MacDiarmid Institute.  This work was supported by project funding from the MacDiarmid Institute and the New Zealand Ministry of Business, Innovation \& Employment Magnetic Devices contract RTVU1203. 
\\

\bibliography{STNO_Project-CMG_PMA_Paper}

\end{document}